\documentclass[twocolumn]{aastex631}

\shorttitle{Dynamical Environment within the Habitable Zone}
\shortauthors{Stephen R. Kane}

\begin{document}

\title{The Dynamical Environment within the Habitable Zone of the
  Gaia-4 and Gaia-5 Planetary Systems}

\author[0000-0002-7084-0529]{Stephen R. Kane}
\affiliation{Department of Earth and Planetary Sciences, University of
  California, Riverside, CA 92521, USA}
\email{skane@ucr.edu}


\begin{abstract}

Exoplanetary systems exhibit a broad range of architectures which, in
turn, enable a variety of dynamical environments. Many of the known
planetary systems do not transit the host star, and so we measure the
minimum masses of their planets, making it difficult to fully assess
the dynamical environment within the system. Astrometry can resolve
the mass ambiguity and thus allow a more complete dynamical analysis
of systems to be conducted. Gaia-4 and Gaia-5 are two such systems,
whose study with radial velocities and data from the Gaia mission
revealed that each star harbors a massive planet on a highly eccentric
orbit. In this work, we provide the results of a dynamical analysis of
each system, including calculations of the Habitable Zone (HZ), from
which we show that the presence of the known companions largely
exclude the presence of planets within the HZ. We discuss the
diagnostics of potential past planet-planet scattering events, and the
occurrence of similar systems whereby a giant planet on an eccentric
orbit can substantially disrupt orbital integrity of terrestrial
planets. These ``wrecking ball'' systems have an impact on the target
selection for planned direct imaging missions that seek to identify
potentially habitable environments.

\end{abstract}



\section{Introduction}
\label{intro}

The detection of thousands of exoplanets has revealed a broad range of
planetary configurations
\citep{ford2014,winn2015,mishra2023a,mishra2023b}. This extraordinary
diversity of discovered planetary system architectures has enabled the
emergence of statistical studies into their prevalence relative to the
solar system \citep{martin2015b,horner2020b,raymond2020a,kane2021d}
and provided insight into planetary formation processes
\citep{morbidelli2007b,raymond2008b,raymond2009b,kane2023a}. Radial
velocity (RV) surveys continue to be a strong source of multi-planet
system discoveries, including Keplerian orbital parameters
\citep{fischer2016,butler2017,reiners2018b,rosenthal2021}. Although
the RV technique excels at determining planetary orbits, it is only
the minimum planetary masses that are calculated from the measurable
quantities due to the unknown inclination of the orbits.  The true
mass of the planet may be constrained through a combination of RV and
direct imaging \citep{kane2019b,dalba2021b}, but a formidable
combination for revealing the orbital inclination and planet mass is
via combining RVs with astrometry
\citep{wright2009b,brandt2019,winn2022,yahalomi2023,feng2025}. Since
non-transiting planets comprise the overwhelming majority of those yet
to be discovered, the unlocking of true planet masses will enable a
significant improvement in understanding the architectures and
dynamics of planetary systems. Such knowledge will further allow a
more complete assessment for the dynamical viability of potential
terrestrial planets within the Habitable Zone (HZ) of their systems
\citep{kasting1993a,kane2012a,kopparapu2013a,kopparapu2014,kane2016c,hill2018,hill2023}.

An important step toward the astrometric detection of exoplanets has
been facilitated through the data releases from the Gaia mission
\citep{perryman2014c,brandt2018,brandt2021a,brown2021}. These data
have been used to refine stellar and planetary properties
\citep{stassun2017,berger2018c,fulton2018b,delaverny2025} and to
demonstrate that some previously detected RV companions do not lie in
the planetary mass regime \citep{keifer2019d}. Two recent discoveries
showed the power of combining Gaia astrometry with RVs through the
announcement of Gaia-4b and Gaia-5b: two massive sub-stellar
companions orbiting low-mass stars \citep{stefansson2025}. The
combined data allowed the measurement of the true masses of the
companions, which are 11.80 and 20.87 Jupiter masses ($M_J$) for
Gaia-4b and Gaia-5b, respectively, placing the latter near the brown
dwarf boundary. Furthermore, both companions are in wide-separation,
eccentric orbits relative to the broader eccentricity exoplanet
distribution \citep{shen2008c,kane2012d,vaneylen2015}, situating them
within the important demographic of ``wrecking ball'' planets, which
are those that can effectively gravitationally dominate the dynamics
of the majority of the system \citep{kane2019e}. Such systems can
serve as RV benchmarks, as the RV signal is less likely to be
contaminated by the presence of as-yet undiscovered planets that lie
beneath the detection threshold \citep{brewer2020}. However, these
systems also have dramatically reduced habitability prospects due to
the dynamical disruption of stable orbits within the HZ
\citep{kopparapu2010,kane2012e,kane2024d,kane2024e}. The Gaia-4 and
Gaia-5 systems thus have great importance in their companions having
true mass measurements allowing a robust dynamical assessment, their
relevance to exoplanet demographics, and the implications for direct
imaging target selection \citep{laliotis2023,harada2024b,tuchow2024}.

In this paper, we present the results of a dynamical analysis of the
Gaia-4 and Gaia-5 systems in relation to the HZ within each system,
taking advantage of the stellar and planetary measurements provided by
astrometric and RV data. Section~\ref{arch} describes the architecture
for each system, and the calculations of the HZ boundaries. The
description of the methodology used for the dynamical analysis is
provided in Section~\ref{stab}, along with the results of the
simulations. In Section~\ref{discussion} we discuss the potential for
past planet-planet interactions, the occurrence of wrecking ball
scenarios, and the relevance to direct imaging missions. We provide a
summary of our work and concluding remarks in
Section~\ref{conclusions}.


\section{System Architectures and Habitable Zones}
\label{arch}

Each of the planetary systems considered here contain a central body
that consists of a main sequence star (Gaia-4 is a K dwarf and Gaia-5
is a M dwarf), orbited by a giant planet on an eccentric orbit. We
adopt the stellar and planetary parameters provided by
\citet{stefansson2025}, also shown in Table~\ref{tab:params}. We
include those stellar parameters most relevant to our analysis,
including stellar mass ($M_\star$), stellar radius ($R_\star$),
effective temperature ($T_\mathrm{eff}$), luminosity ($L_\star$), and
distance ($d$). The planet parameters include the planet mass ($M_p$),
orbital period ($P$), eccentricity ($e$), argument of periastron
($\omega$), and orbital inclination ($i$). \citet{stefansson2025} does
not provide the semi-major axis ($a$), but we calculate those as
follows: $a = 1.1705$~AU for Gaia-4b and $a = 0.7021$~AU for Gaia-5b.

\begin{deluxetable}{lrr}
\tablecaption{\label{tab:params} Planetary system parameters.}
\tablehead{
  \colhead{Parameter} & 
  \colhead{Gaia-4} & 
  \colhead{Gaia-5}
}
\startdata
\sidehead{\bf{Stellar Parameters}}
  $M_\star$ ($M_\odot$) & 0.644  & 0.339 \\
  $R_\star$ ($R_\odot$) & 0.624  & 0.345 \\
  $T_\mathrm{eff}$ (K)  & 4034   & 3447  \\
  $L_\star$ ($L_\odot$) & 0.100  & 0.015 \\
  $d$ (pc)              & 73.7   & 41.248 \\
  Inner OHZ (AU)        & 0.256  & 0.100 \\
  Inner CHZ (AU)        & 0.325  & 0.127 \\
  Outer CHZ (AU)        & 0.614  & 0.246 \\
  Outer OHZ (AU)        & 0.648  & 0.260 \\
\hline
\sidehead{\bf{Planet Parameters}}
  $M_p$ ($M_J$)         & 11.80 & 20.87   \\
  $P$ (days)            & 571.3 & 358.62  \\
  $a$ (AU)              & 1.1705 & 0.7021 \\
  $e$                   & 0.338  & 0.6423 \\
  $\omega$ ($\degr$)    & 180.3  & 271.54 \\
  $i$ ($\degr$)         & 116.9  & 129.7  \\
\enddata
\end{deluxetable}

\begin{figure*}
    \begin{center}
        \begin{tabular}{cc}
            \includegraphics[angle=270,width=8.5cm]{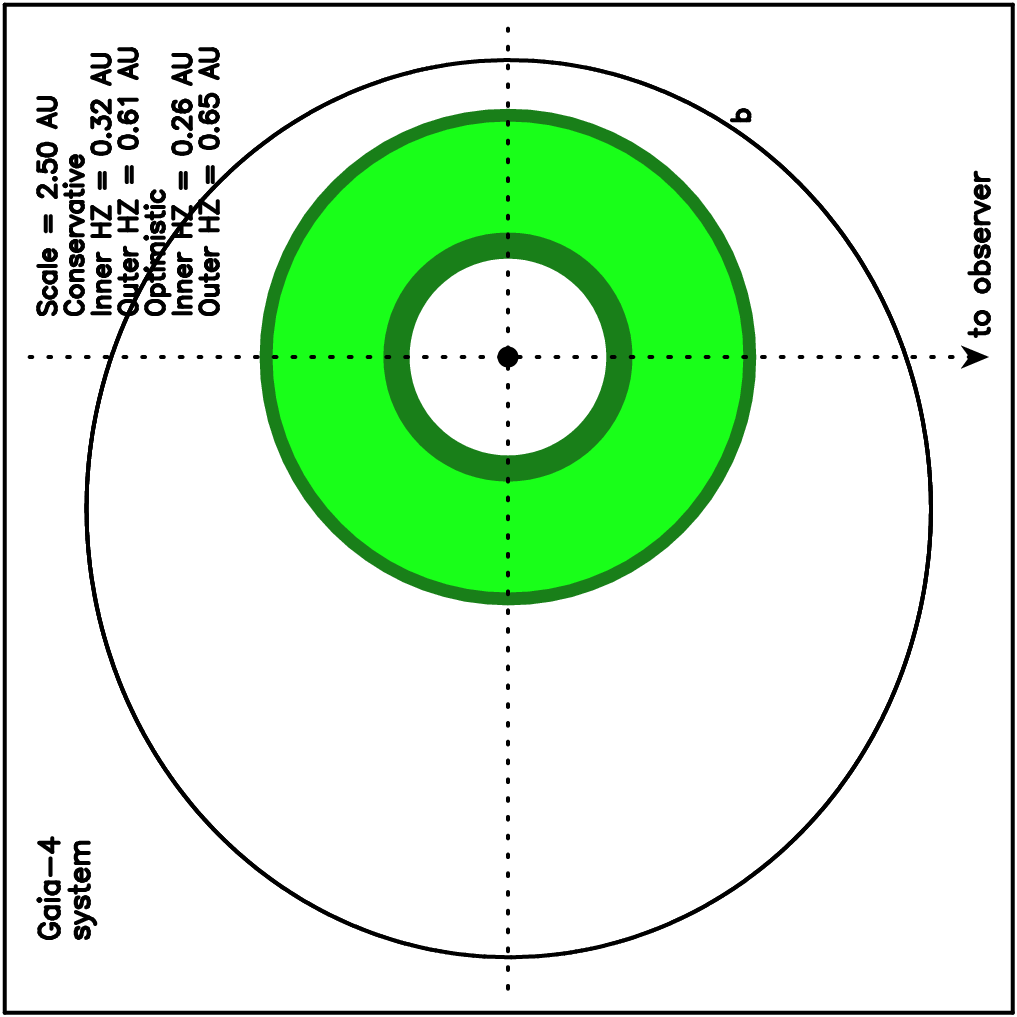} &
            \includegraphics[angle=270,width=8.5cm]{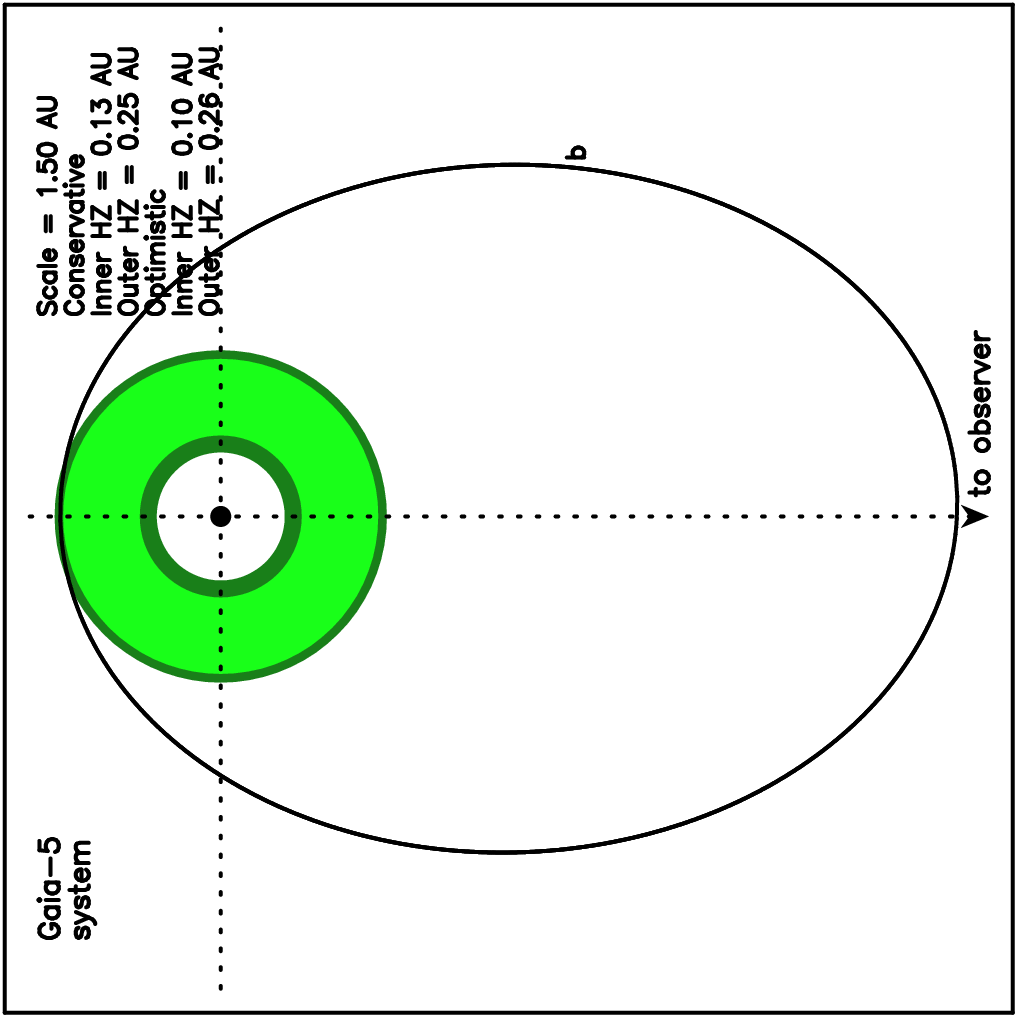} \\
        \end{tabular}
    \end{center}
  \caption{System architectures and HZ boundaries for the Gaia-4
    (left) and Gaia-5 (right) systems, showing the orbit of the planet
    in each case. The HZ regions are shown in green, where light green
    and dark green indicate the CHZ and OHZ, respectively. The scale
    of the figure panels are 2.5~AU and 1.5~AU along each side for
    Gaia-4 and Gaia-5, respectively.}
  \label{fig:hz}
\end{figure*}

The calculation of the HZ for each system is sensitive to the
precision of the stellar parameters \citep{kane2014a}. We used the
stellar properties provided by \citet{stefansson2025} (see
Table~\ref{tab:params}), adopting the methodology of
\citet{kopparapu2013a,kopparapu2014}. These calculations resulted in
two primary sets of boundaries: the conservative HZ (CHZ), based on 1D
atmospheric models of runaway greenhouse and maximum greenhouse
transitions, and the optimistic HZ (OHZ), based on empirical evidence
that Venus and Mars may have hosted surface liquid water in their
pasts. A full description of the CHZ and OHZ boundaries may be found
in \citet{kane2016c}. For Gaia-4, the HZ regions span 0.325--0.614~AU
and 0.256--0.648~AU for the CHZ and OHZ, respectively. For Gaia-5, the
HZ regions span 0.127--0.246~AU and 0.100--0.260~AU for the CHZ and
OHZ, respectively. These HZ boundary values are also provided in
Table~\ref{tab:params}.

Shown in Figure~\ref{fig:hz} are the system architectures for each of
the systems: Gaia-4 (left panel) and Gaia-5 (right panel). The extent
of the HZ for each system is shown in green, where light green and
dark green indicate the CHZ and OHZ, respectively. The orbits of the
planets are also indicated as solid lines. Based on the orbital
parameters from Table~\ref{tab:params}, the periastron distances of
the planets are 0.755~AU and 0.251~AU for Gaia-4b and Gaia-5b,
respectively. For Gaia-4b, the periastron distance brings it close to
the HZ, but it does not quite enter that region. For Gaia-5b, the
planet enters the outer OHZ region and spends 2.3\% of its orbit
within the HZ. Particularly given the large relative mass of Gaia-5b,
this entrance into the HZ will have significant consequences for the
stability of other potential planets in HZ, as discussed in later
sections.


\section{Dynamical Stability}
\label{stab}

In this section, we describe the dynamical simulations performed for
each of the systems, and the resulting stability predictions within the
HZ.


\subsection{Dynamical Simulation Methodology}
\label{methods}

The dynamical simulations carried out in this work made use of the
Mercury Integrator Package \citep{chambers1999}, adopting the hybrid
symplectic/Bulirsch-Stoer integrator with a Jacobi coordinate system,
providing more accurate results for multi-planet systems
\citep{wisdom1991,wisdom2006b}. For each of the two systems, we
conducted an injection stability analysis, whereby we inserted an
Earth-mass planet in a circular orbit at a variety of semi-major axes
start locations, similar to the procedure described by
\citet{kane2019c,kane2021a,kane2023c}. The range of semi-major axis
values tested were chosen to encompass the HZ regions described in
Section~\ref{arch} and shown in Table~\ref{tab:params}. Specifically,
we adopted semi-major values for the injected planet in the range
0.20--0.70~AU and 0.08--0.28~AU for the Gaia-4 and Gaia-5 systems,
respectively, in steps of 0.001~AU. These are equivalent to orbital
period ranges of 40.7--266.5~days and 14.2--92.9~days for the Gaia-4
and Gaia-5 systems, respectively. We used a time step of 0.2~days to
provide adequate time sample of perturbations due to the presence of
the known planet, and each simulation was run for $10^7$~years. The
stability outcome for each simulation was determined via an assessment
of the orbital evolution of the injected planet. If the planet does
not survive the full $10^7$~year integration, then that means the
planet was captured by the gravitational well of the host star,
ejected from the system, or subjected to a planet-planet collision.
The survival rate was then calculated for each set of initial starting
conditions.

\begin{figure*}
  \begin{center}
    \includegraphics[angle=270,width=16.0cm]{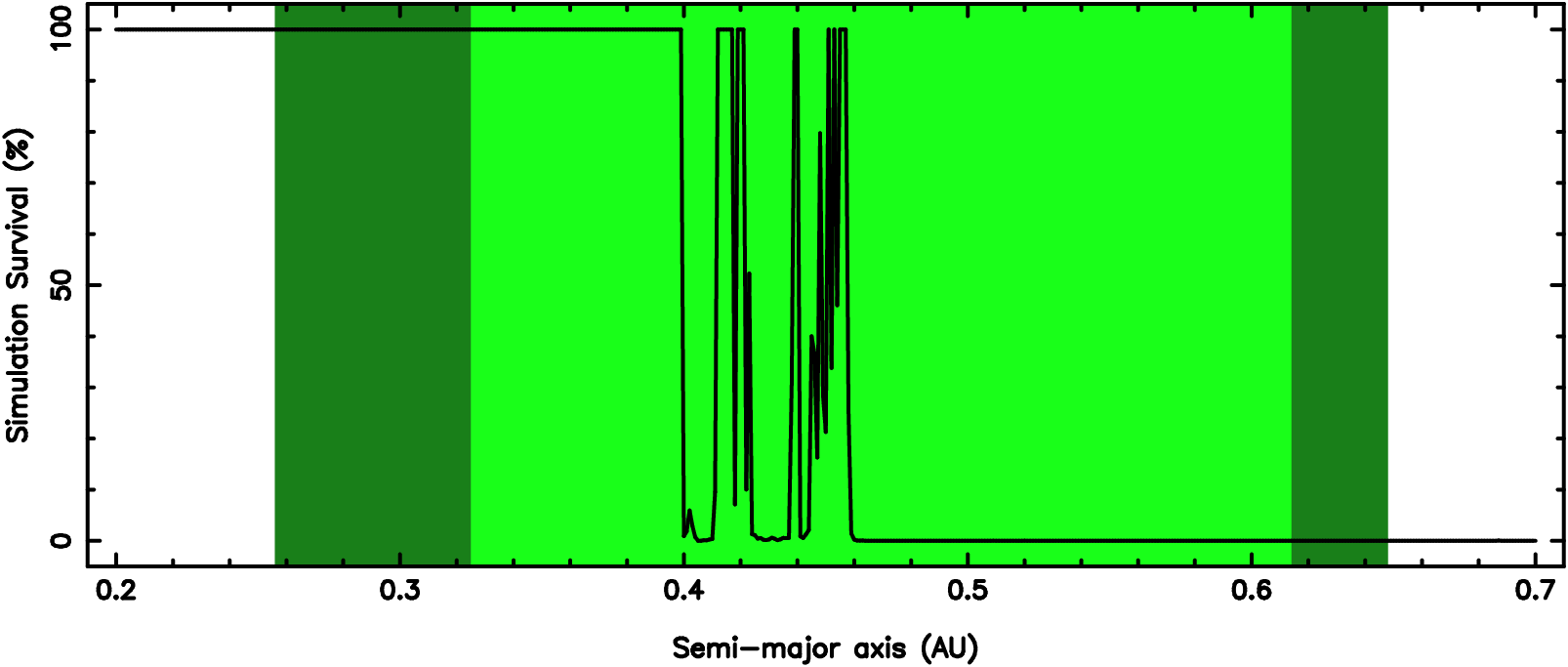} \\
    \includegraphics[angle=270,width=16.0cm]{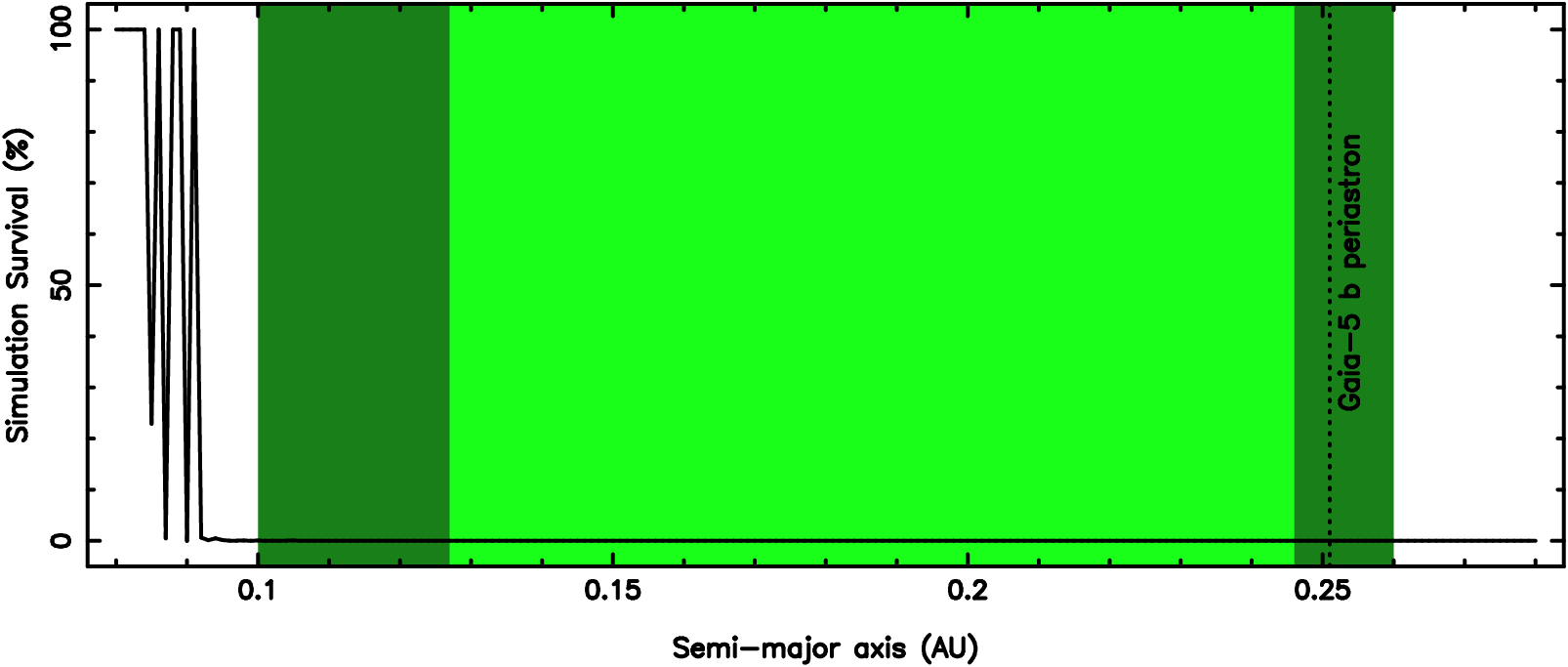}
  \end{center}
  \caption{Percentage of the simulation that the injected Earth-mass
    planet survived as a function of semi-major axis for the Gaia-4
    (top panel) and Gaia-5 (bottom panel) systems. As for
    Figure~\ref{fig:hz}, the CHZ is shown in light green and the OHZ
    is shown in dark green. The periastron passage of Gaia-5b is
    indicated by the vertical dotted line.}
  \label{fig:sim}
\end{figure*}


\subsection{Stability Within the Habitable Zone}
\label{hz}

As noted in Section~\ref{arch} the known companions for each of the
systems have periastron locations that approach the HZ outer boundary,
and indeed enter the OHZ in the case of Gaia-5b. As described in
Section~\ref{methods}, we tested the viability of additional planetary
orbits within the HZ via a suite of Earth-mass planet injections,
whose semi-major axes encompass the HZ within each system. The
percentage survival rates at each of the injected locations were then
calculated at each location. These results are shown in
Figure~\ref{fig:sim}, where the Gaia-4 calculations are represented by
the solid line in the top panel, and similarly for the Gaia-5
calculations in the bottom panel. The HZ is depicted using the same
color scheme as for Figure~\ref{fig:hz}, where light green and dark
green indicate the CHZ and OHZ, respectively.

These results show that more than half of the HZ in the Gaia-4 system
is rendered unstable by the presence of the known massive
companion. To understand the islands of stability visible in the
semi-major axis range 0.40--0.46~AU, we calculated the locations of
mean motion resonance (MMR). Such MMR locations have special
importance for stability considerations within planetary systems,
including the solar system, particularly as secular perturbations may
lead to stable/unstable islands depending on the orbital architectures
\citep{peale1976a,beauge2003b,petrovich2013,goldreich2014,hadden2019b}. In
the case of Gaia-4, the 5:7 and 2:3 MMR locations occur at 0.423~AU
and 0.443~AU, respectively. These MMR locations align well with the
stability islands and thus may be the primary contributor toward those
high simulation survival rates.

The results for the dynamical simulation for Gaia-5, shown in the
bottom panel of Figure~\ref{fig:sim}, suggest a far more severe
outcome than for Gaia-4. The vertical dotted line indicates the
periastron location for the companion. This orbital penetration into
the HZ, combined with the relatively high mass of the companion,
effectively rule out all stable orbits within the HZ of the system.


\section{Discussion}
\label{discussion}


\subsection{Angular Momentum Deficit}
\label{amd}

The high eccentricity of the massive planets within the Gaia-4 and
Gaia-5 systems may be be signatures of significant past interactions,
such as disk interactions during formation \citep{clement2021e} or a
potentially dynamically turbulent past regarding planet-planet
scattering events
\citep{chatterjee2008,ford2008c,kane2014b,carrera2019b,childs2025}. A
method to diagnose the dynamical history is the use of the angular
momentum deficit (AMD) for the system \citep{laskar1997}. The AMD
provides a calculation of the difference in total angular momentum
between the eccentric orbits present within a system and equivalent
circular orbits, possibly the result planet ejection scenarios
\citep{laskar2017,he2020c}.

\begin{figure}
  \includegraphics[angle=270,width=8.5cm]{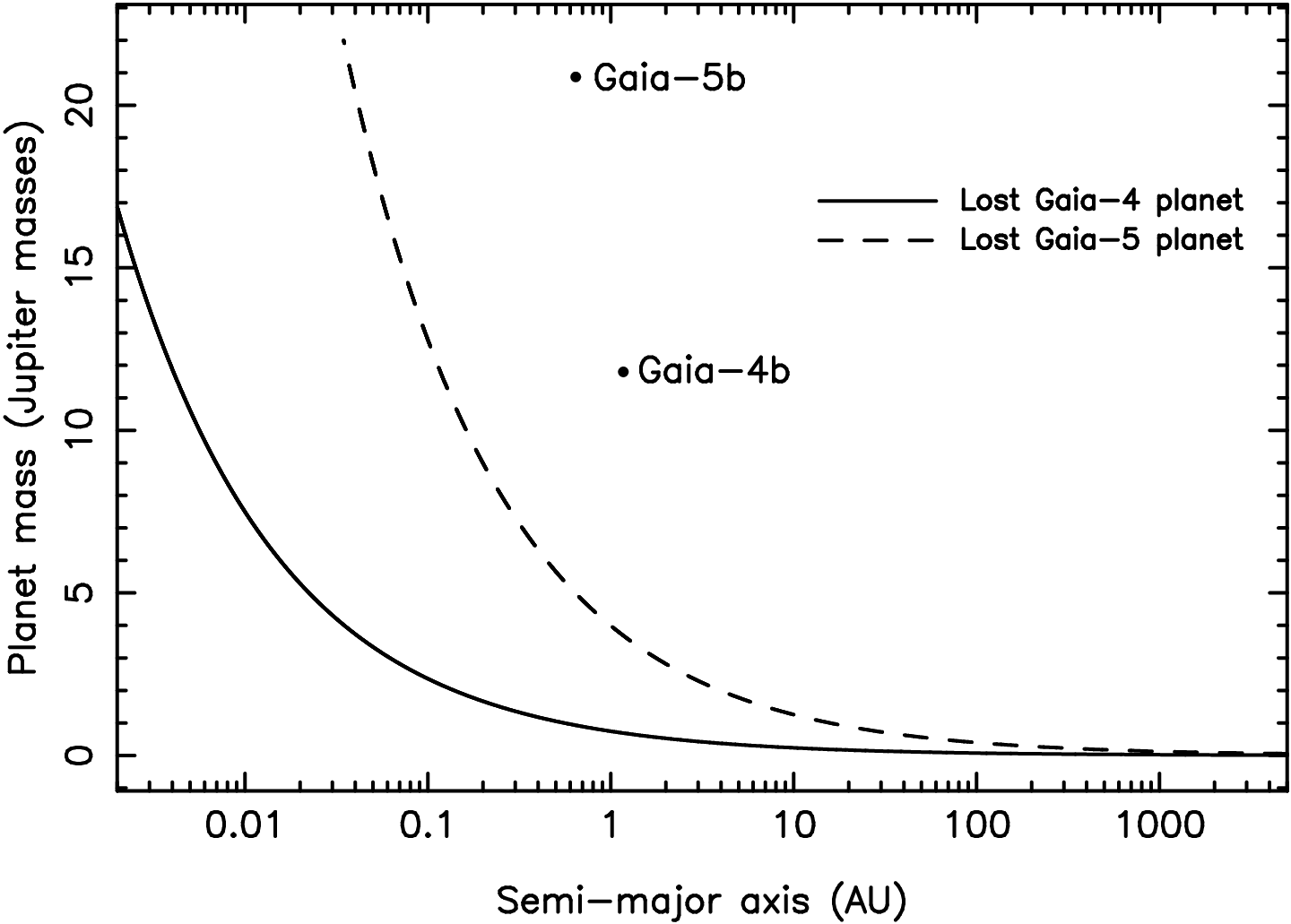}
  \caption{The mass and semi-major axis of a possible lost planet in a
    circular orbit whose angular momentum equals the angular momentum
    deficit (AMD) for each of the considered systems: Gaia-4 (solid
    line) and Gaia-5 (dashed line). The dots indicate the planet mass
    and semi-major axis of the known companions.}
  \label{fig:amd}
\end{figure}

For Gaia-4 and Gaia-5, we calculate AMD values of $5.06 \times
10^{42}$~kg\,m$^2$/s and $1.96 \times 10^{43}$~kg\,m$^2$/s,
respectively, the latter of which is approximately equivalent to the
angular momentum of Jupiter in the solar system. A given AMD value is
degenerate with the planetary mass and semi-major axis. We thus
calculated the planet mass and semi-major axis values that produce the
AMD for both of the systems. These calculations are represented in
Figure~\ref{fig:amd}, where the solid curves and dashed curves
indicate the calculations for the Gaia-4 and Gaia-5 systems,
respectively. We also plot the locations of Gaia-4b and Gaia-5b for
comparison. As Figure~\ref{fig:amd} shows, the possible properties of
a potentially lost planet are extremely broad, and also assumes that
there was only one other planet involved in a possible scattering
event. Note also that the small semi-major axis range ($\lesssim
0.01$~AU) will be unphysical as this approaches the Roche limit or
even the star itself ($R_\odot \sim 0.005$~AU).


\subsection{The Occurrence of Habitable Zone Wrecking Balls}
\label{occurrence}

\begin{figure*}
    \begin{center}
        \begin{tabular}{cc}
            \includegraphics[angle=270,width=8.5cm]{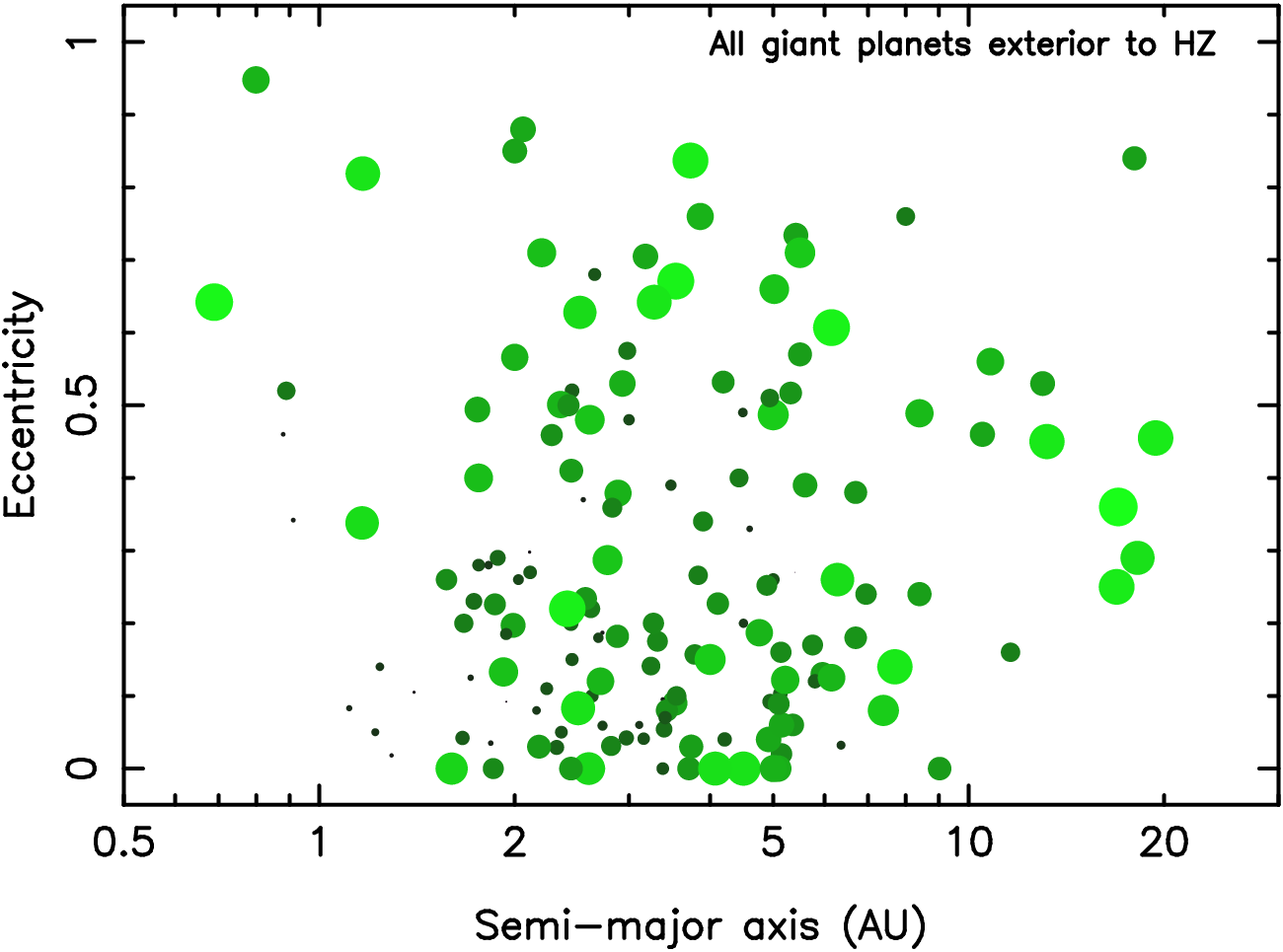} &
            \includegraphics[angle=270,width=8.5cm]{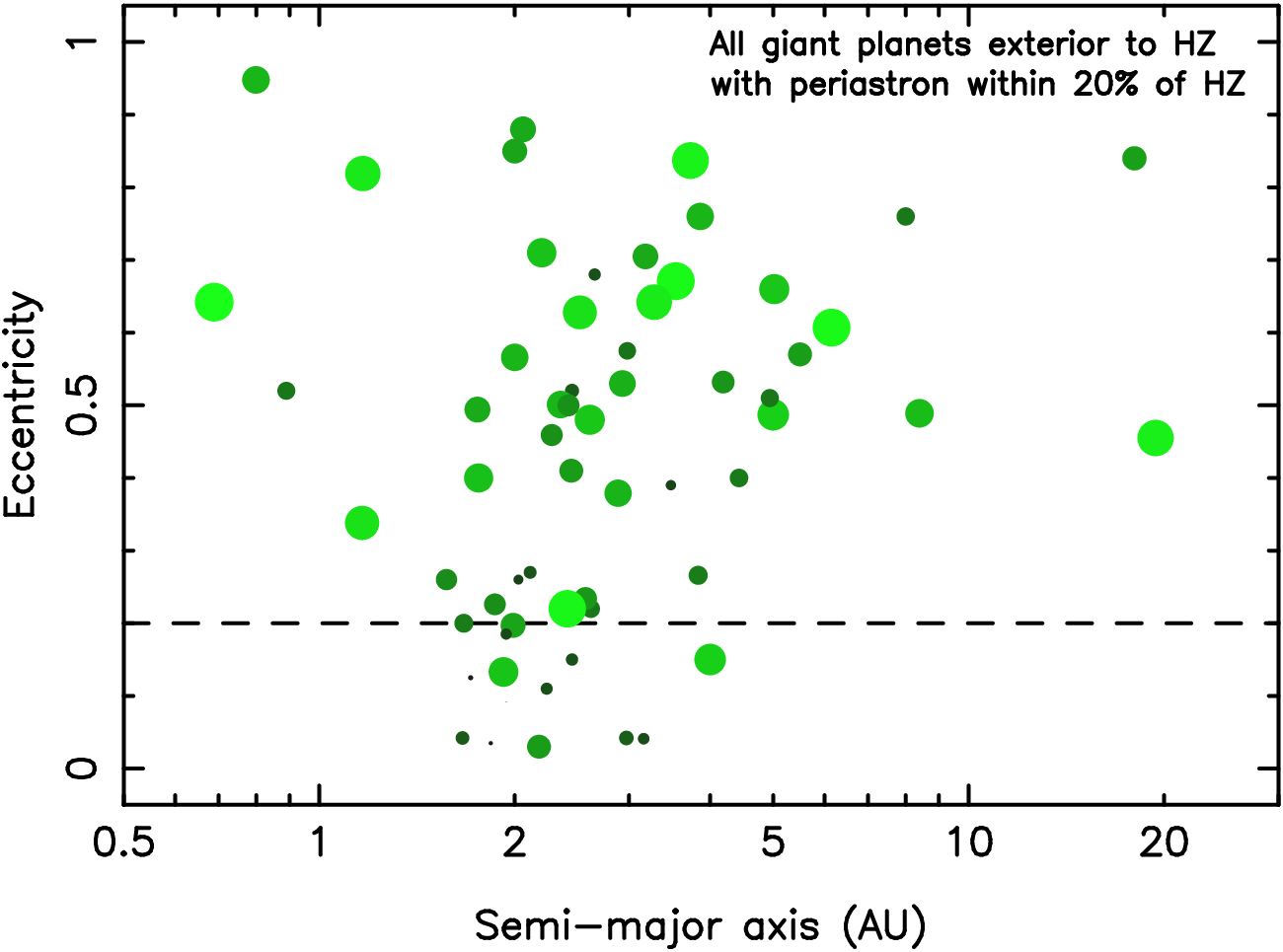} \\
        \end{tabular}
    \end{center}
  \caption{The eccentricity distribution of giant planets ($M_p >
    0.1$~$M_J$) whose semi-major axis lies beyond the OHZ outer
    boundary (left panel). Both the shade and the size of the plotted
    data are logarithmically proportional to the planet mass, where
    dark green indicates a low mass and light green indicates a high
    mass. The right panel shows a subset (the wrecking ball planets)
    whose orbits either pass through the HZ or whose periastron
    distance lies within 20\% of the outer OHZ boundary.}
  \label{fig:occ}
\end{figure*}

Long duration exoplanet surveys have found that giant planets beyond
the snow line are relatively rare, even for solar-type stars
\citep{wittenmyer2011a,wittenmyer2016c,wittenmyer2020b,fulton2021,rosenthal2021,bonomo2023}. These
cold giant planets play important roles in shaping the architecture of
planetary systems, and potential volatile delivery to the inner
regions of the system
\citep{obrien2014a,raymond2017b,venturini2020b,kane2024a,kane2025a}. For
the special case of wrecking ball planets, we investigated their
occurrence relative to the HZ through the use of data from the
Habitable Zone Gallery \citep{kane2012a}, which are current as of July
4, 2025. A total of 5143 planets formed the initial dataset, from
which we selected those whose mass is $>0.1$~$M_J$ and whose
semi-major axis lies beyond the outer edge of the OHZ. These criteria
formed a sample of 162 planets, and are represented in the left panel
of Figure~\ref{fig:occ}. Both the shade and the size of the plotted
data are logarithmically proportional to the planet mass. The median
planet mass for those with eccentricities of $e > 0.2$ is 2.71~$M_J$,
and the median planet mass for $e < 0.2$ is 1.92~$M_J$. Thus, this
cold giant planet population exhibit the known correlation of
eccentricity with planet mass \citep{ribas2007}, although this may be
partially the result of observational bias
\citep{kane2007a,zakamska2011,wittenmyer2019a}. The planet population
shown in the left panel of Figure~\ref{fig:occ} does not necessarily
interact with the HZ however, and so we created a subset of the planet
population whose orbits either pass through the HZ or whose periastron
distance lies within 20\% of the outer OHZ boundary. That subset
population is shown in the right panel of Figure~\ref{fig:occ} and
contains 62 planets, 48 of which have eccentricities $e > 0.2$
(indicated by the horizontal dashed lines), the latter of which we
designate as the wrecking ball population. The wrecking ball planets
thus comprise 30\% of the total cold giant planet population shown in
the left panel of Figure~\ref{fig:occ}. As earlier described, these
wrecking ball planets play an important role in determining the
occurrence of potentially habitable
planets\citep[e.g.,][]{dressing2013,kopparapu2013b,kunimoto2020b,bryson2021},
and can compromise the viability of direct imaging targets that aim to
detect and characterize habitable worlds \citep{kane2024d,kane2024e}.


\subsection{Suitability as Direct Imaging Targets}
\label{imaging}

\begin{figure}
  \includegraphics[angle=270,width=8.5cm]{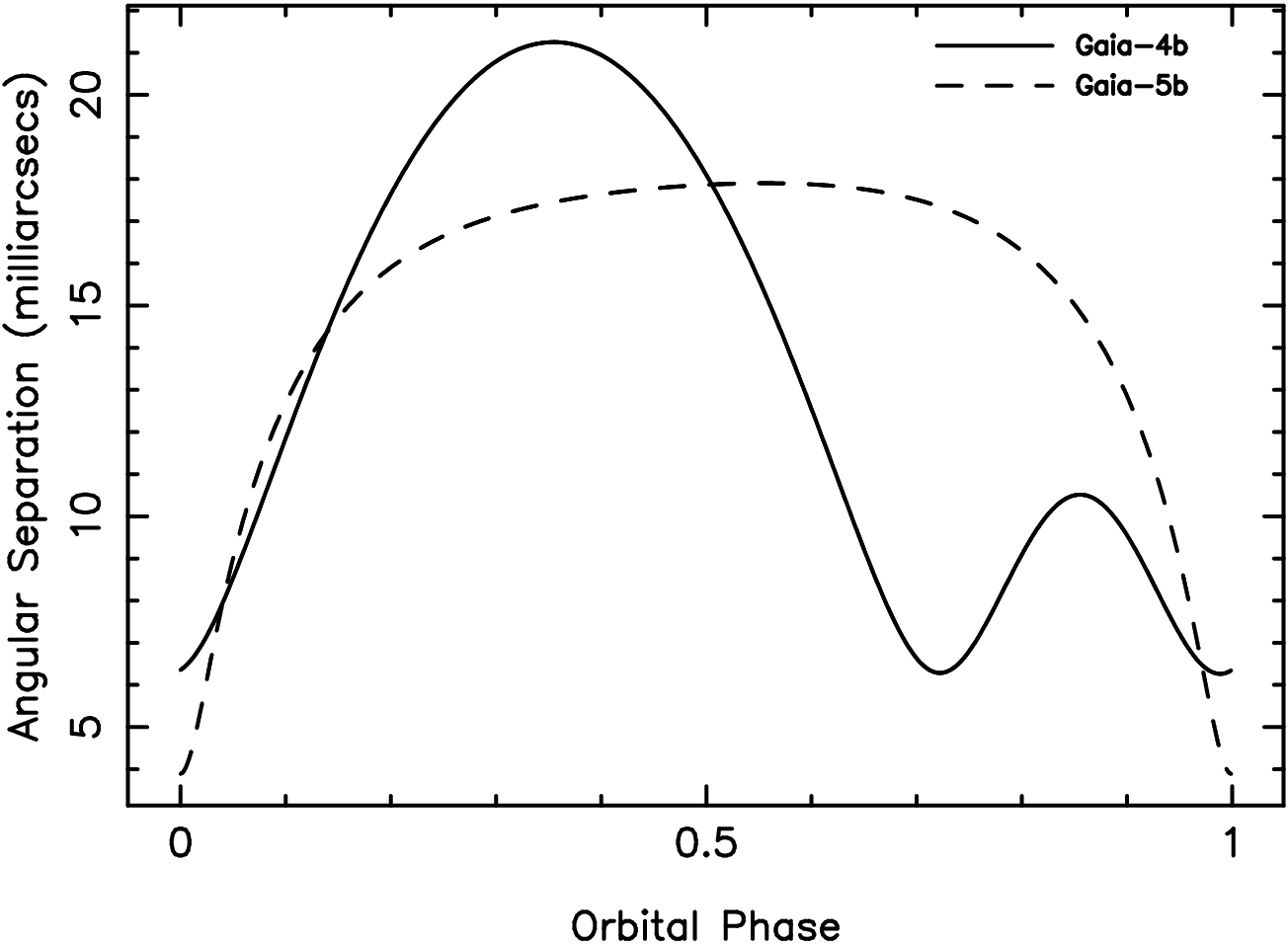}
  \caption{Angular separation of Gaia-4b (solid line) and Gaia-5b
    (dashed line) from their respective host stars as a function of
    their orbital phase.}
  \label{fig:imaging}
\end{figure}

Although the HZ of the systems are largely unviable for terrestrial
planetary orbits, the astrometric detection of Gaia-4b and Gaia-5b,
along with their wide separation and eccentric orbits, make the
companions themselves potentially interesting as direct imaging
targets. A space-based telescope equipped with a coronagraph and/or
starshade would likely provide the greatest capability for directly
imaging the companions. For example, the Nancy Grace Roman Space
Telescope equipped with a Coronagraph Instrument (CGI) would achieve
contrast ratio detection limits of $10^{-8}$ at angular separations as
low as 150~milliarcsecs (mas) for bright ($V < 6$) stars
\citep{turnbull2021}. Furthermore, an idealized coronagraph operating
at 1~$\mu$m on an 8~m telescope would achieve an inner working angle
of 26~mas \citep{tuchow2024}.

The orbital parameters provided in Table~\ref{tab:params}, including
the orbital inclinations, allow the calculation of the angular
separation of the companions from their respective host stars as a
function of their orbital phase. These calculations are shown in
Figure~\ref{fig:imaging}, where the angular separations for Gaia-4b
and Gaia-5b are represented by the solid and dashed lines,
respectively. The angular separations of the companions span a wide
range due to their orbital eccentricities, dropping below
10~mas. However, most of the orbital phase is spent near apastron at
separations $> 15$~mas, increasing the opportunities for observation
\citep{kane2013c,kane2018c}. Though their maximum angular separations
remain challenging for present facilities, such wide orbit companions
may be ideal science cases for proposed future facilities, such as the
Habitable Worlds Observatory \citep{harada2024b}.


\section{Conclusions}
\label{conclusions}

The dynamical environment within exoplanetary systems plays a crucial
role in shaping the observed distribution of planetary architectures,
and also can significantly impact the potential for habitable
terrestrial planets. The synergistic use of RV and astrometric
techniques provides an important avenue to more accurately assess the
dynamical consequences of giant planets, especially those that lie on
eccentric orbits that may compromise the stability of planets within
the HZ. The Gaia-4 and Gaia-5 systems are both excellent examples of
wrecking ball systems where the true mass of the companions have been
extracted, along with their orbital elements. The research presented
here shows that Gaia-4b excludes planetary orbits for the outer half
of the HZ, with stable possibilities at the HZ inner half, including
locations of MMR. For Gaia-5, the relatively high mass and
eccentricity of the companion rules out any planetary occupancy of the
HZ.

Wrecking ball systems are an important part of the exoplanet
demographic for several reasons. First, they are an essential part of
the planetary system formation and evolution story, which may be
indicative of past planet-planet scattering events. Second, they can
result in planetary systems being rendered unsuitable as targets for
missions that aim to detect and characterize potentially habitable
worlds. Since the vast majority of planets do not transit their host
star, the combination of RV and astrometry form an increasingly
critical pathway for a full census of system architectures, and their
consequences for long-term habitability prospects within each system.


\section*{Acknowledgements}

The author would like to thank the anonymous referee and Gudmundur
Stef\'ansson for insightful feedback on the manuscript. This research
has made use of the Habitable Zone Gallery at hzgallery.org. The
results reported herein benefited from collaborations and/or
information exchange within NASA's Nexus for Exoplanet System Science
(NExSS) research coordination network sponsored by NASA's Science
Mission Directorate.


\software{Mercury \citep{chambers1999}}




\end{document}